\documentclass[%
reprint,
showpacs,preprintnumbers,
 amsmath,amssymb,
 aps,
prl,
]{revtex4-1}

\usepackage{graphicx}
\usepackage{dcolumn}
\usepackage{bm}

\begin{document}


\title{Propagation of stationary exotermic transition front with nonstationary oscillatory tail}


\author{V.V. Smirnov$^1$}
\email{vvs@polymer.chph.ras.ru}
\author{O.V.Gendelman$^2$}
\email{ovgend@tx.technion.ac.il}
\author{L.I. Manevitch$^1$}

\affiliation{$^1$
 Institute of Chemical Physics, RAS, Moscow, Russia\\
119991, 4 Kosygin str., Moscow, Russia
}
\affiliation{$^2$ Faculty of Mechanical Engineering, Technion-Israel Institute of Technology, Haifa, 32000, Israel}

\date{\today}

\begin{abstract}
We consider a propagation of exotermic transition front in a discrete conservative oscillatory chain. Adequate description of such fronts is a key point in prediction of important transient phenomena, including phase transitions and topochemical reactions. Due to constant energy supply, the transition front can propagate with high velocities, precluding any continuum-based considerations. Stationary propagation of the  front is accompanied by formation of a non-stationary oscillatory tail with complicated internal structure. We demonstrate that the structure of the oscillatory tail is related to a relationship between phase and group velocities of the oscillations. We suggest also an approximate analytic procedure, which allows one to determine all basic characteristics of the propagation process: velocity and width of the front, frequency and amplitude of the after-front oscillations, as well as the structure of the oscillatory tail. As an example, we consider a simple case of biharmonic double-well on-site potential. Numeric results nicely conform to the analytic predictions.

\end{abstract}

\maketitle



Structural and conformational transitions in solid state, as well as some chemical transformations, like topochemical reactions, occur through a propagation of the boundary, which divides the regions belonging to different stationary (stable or metastable) states of the system. At the microscopic level, such systems are frequently modeled with the help of low – dimensional lattices or even atomic chains. The possibility of two (or more) stable ground states is taken into account through introduction of double- or multiple-well on-site potential. Consequently, the transition propagation corresponds to a motion of localized elementary excitations, such as topological solitons (kinks) \cite{Scott, Peyrard, Braun, Krumhansl, Heeger, Manevitch08}.

In vast majority of such models, the on-site potential has energetically degenerate minima. This assumption turns out to be crucial for analysis of the dynamics; in continuum approximation, the topological soliton corresponds to a separatrix in the state space of the system. If the potential minima have different energies, such heteroclinic solution, which transfers the system from one minimum to the other, just does not exist in the continuum limit.  However, in many realistic systems, the energies of the ground states are not equal, and one still observes the stationary transition fronts. Traditionally, stationary transitions in the non-degenerate systems are treated with the help of semi-phenomenological models. The latter are based on thermodynamical variables, heat reservoirs and negative feedback \cite{Scott, Bakanas03, Carpio03, Carpio04}. The other popular approach introduces artificial viscous damping; then, the cases of strong and weak damping are considered \cite{Braun, Peyrard84}
In such models, stationary solutions of the required type are indeed available. Still, there is no physical basis to introduce the dissipative forces at the atomic scale. Even in the semi-phenomenological models, it might be problematic to use the thermodynamical variables, since the transition front can move with very high (even close to sonic) velocity. In the same time, thermalization may occur at substantially longer time scale. The other difficulty lies in the fact that the transition zone may be extremely narrow, just of a few interatomic distances. So, the use of continuum approximation in such cases is at least doubtful. In our treatment we deal with conservative model. Moreover, energy conservation is explicitly used in the analysis.

In certain one-dimensional Hamiltonian chains with complicated structure \cite{Manevitch92,  Manevitch94, Manevitch2001, Manevitch08} one can observe and describe the transition from the initial state with relatively high energy to certain "intermittent" state. Then, the "intermittent" state disintegrates far enough from the front, and the system is attracted to a final low-energy state. Therefore, the transition comprises two well-distinguished stages and the first stage can be described in terms of the traditional continuum approximation.
However, the solution with the "intermittent" quasi-degenerate state is by no means generic and is observed only in very special models with a multi-atomic elementary cell or in monoatomic lattices with complicated interplay of the nearest- and next-neighbor interactions \cite{Manevitch92, SigalovPRE, Manevitch08}. In simple and more generic models, other patterns of the front propagation are observed. Namely, an oscillatory zone is formed immediately after the transition front.This oscillatory zone comprises primarily the oscillations with a single frequency. Such behavior seems natural since in order to ensure the stationary front propagation one needs the oscillatory field with unique phase velocity equal to the front speed. For instance, transition regimes of this sort qualitatively describe a process of exotermic collapse in single-wall carbon nanotubes \cite{Chang08}. In the simplest approximation, the collapse dynamics of such nanotube is similar to the dynamics of the monoatomic quasi-one-dimensional chain with the bistable non-degenerate on-site potential (asymmetric sine-Gordon model) \cite{Smirnov11}. Numeric simulations demonstrate that the collapse front is very narrow and propagates with constant velocity. Similar scenarios of the front propagation were mentioned in a number of previous works devoted to martensitic structural transitions \cite{Slepyan01_1, Slepyan01_2, Trusk05}.
In papers \cite{Trusk05} the authors attempted to treat the problem analytically for a simplified model of the linear chain with piecewise parabolic on-site potential (see also \cite{Atkinson65, Flytzanis77}).  In this case the system dynamics is described by linear equations of motion with appropriate matching conditions in the boundary points dividing the parabolic potential wells. One can compute an exact solution that corresponds to the front propagation accompanied by infinite stationary oscillatory tail \cite{Vainchtein10}. This solution seems somewhat problematic in one important respect: it requires stationary oscillatory field (in fact, an additional nonphysical energy source) at infinity; consequently, the predicted values of the front velocity are much higher than obtained in numeric simulations with more natural boundary conditions (say, for free boundaries). Moreover, as it will be shown in the paper, an intimate mechanism governing by the structure of the post-frontal region remained to be un-clarified even for this simple model.

In this Letter, we suggest an approximate approach to treatment of such processes . This approach explains their mechanism and all important characteristics of the observed transition propagation: the velocity and the width of the front, as well as the structure and the frequency spectrum of the oscillatory tail. The linear chain with bi-stable piecewise parabolic potential serves here as a simple benchmark example; however, the method is applicable to the models with more complicated and realistic structure.
Hamiltonian of the atomic chain with linear nearest-neighbor interactions and on-site potential is written as follows:
\begin{equation}
H=\sum_{n} [ \frac{p_{n}^{2}}{2m}+\frac{c^{2}}{2}(\varphi_{n+1}-\varphi_{n})^{2}+U(\varphi_{n}) ]
\label{H1}
\end{equation}

\begin{figure}
a \includegraphics [width=40mm] {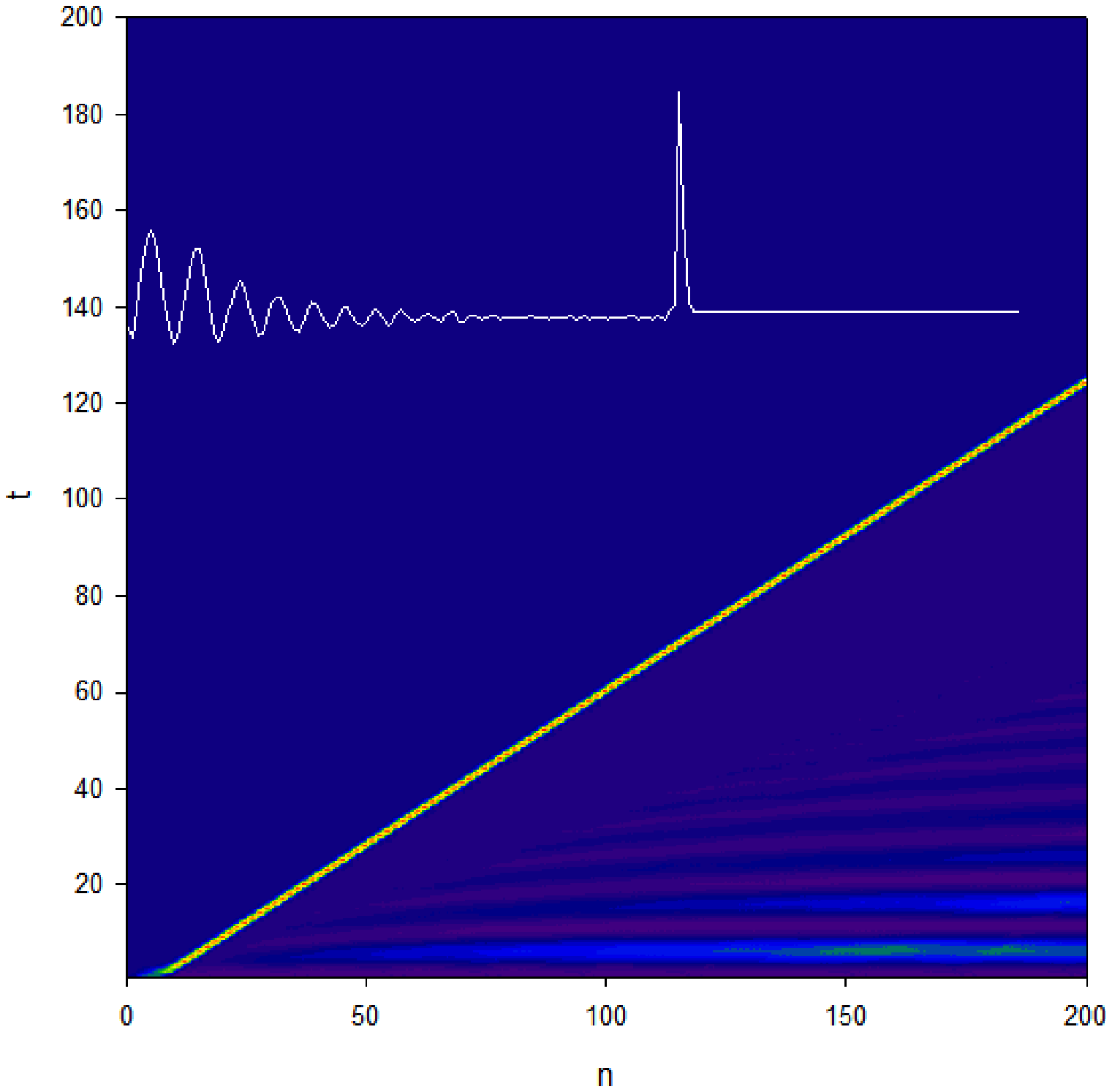}
b \includegraphics [width=40mm] {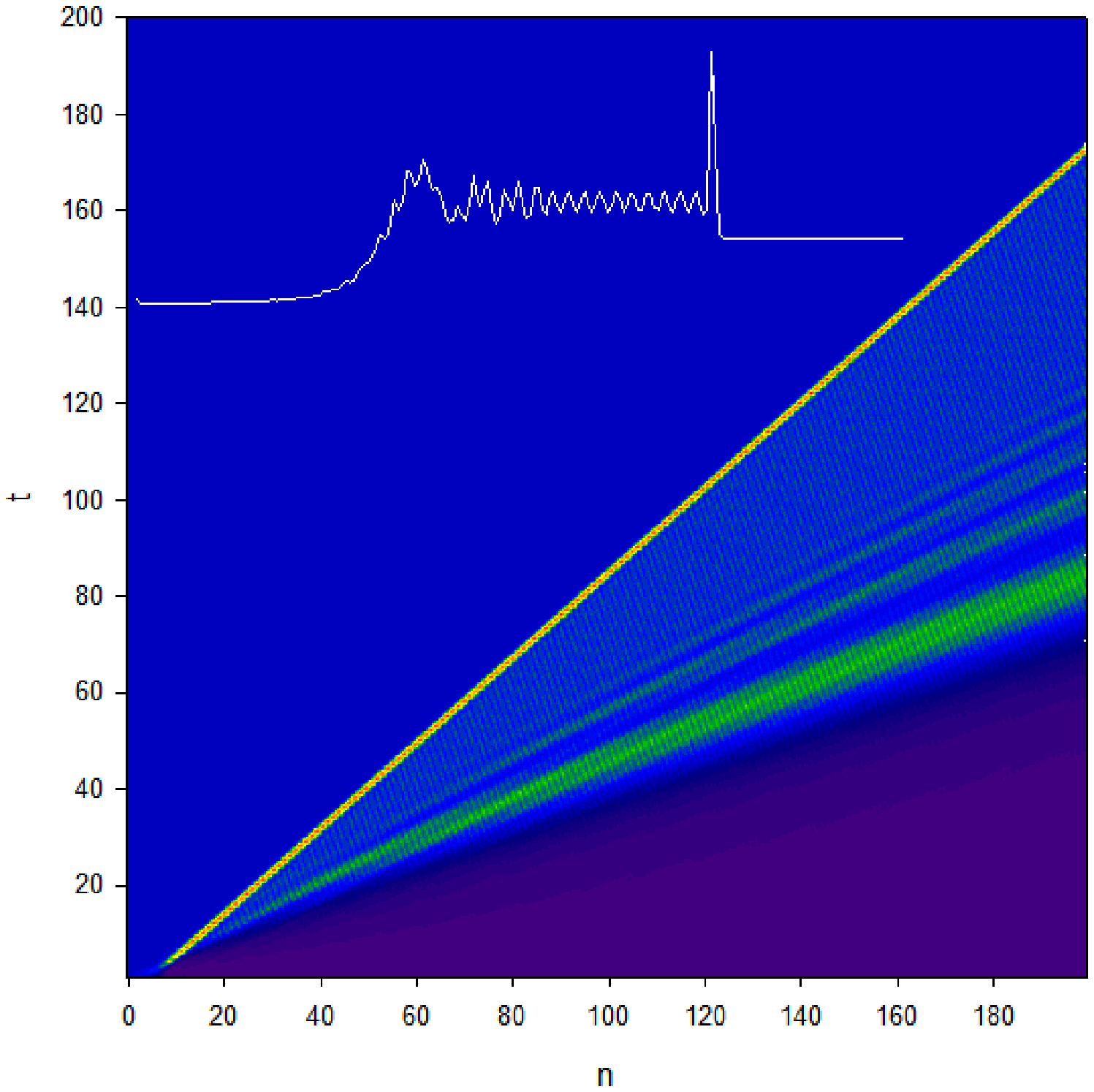}
\caption{Local energy in the system (1-2) with the propagating transition front, $\omega_{0}=\omega_{1}=1$, $c=1$, $E_{b}=0.5$, (a) $Q=0.4$, (b) $Q=1.5$. In the insets we present the instantaneous energy profile of the chain.}
\label{front1}
\end{figure}

Here $\varphi_{n}$ is the displacement of the $n$th particle from the initial equilibrium state; the latter corresponds to the well with higher energy (the metastable state); $p_{n}$ is the particle momentum, $c$ is the stiffness of the nearest-neighbor linear springs and $U(\varphi)$ denotes the non-degenerate on-site potential. Masses of all particles are adopted to be unit. The simplified piecewise parabolic on-site potential is expressed as:
\begin{equation}
\begin{split}
U(\varphi)=
\begin{cases}
	\frac{\omega_{0}^{2}}{2}\varphi^{2}, & \text{if $\varphi\leq \varphi_{b}=\frac{\sqrt{2 E_{b}}}{\omega_{0}}$} \\
	\frac{\omega_{1}^{2}}{2}(\varphi-\varphi_{f})^{2}, & \text{if $\varphi>\varphi_{b}$ }
\end{cases}
\\
\phi_{f}=\frac{\sqrt{2E_{b}}}{\omega_{0}}+\frac{\sqrt{2(E_{b}+Q)}}{\omega_{1}}
\end{split}
\label{onsite}
\end{equation}
Here $\omega_{0}$ and $\omega_{1}$  characterize the well curvatures near the metastable and the stable states, respectively, $\varphi_{b}$ and $\varphi_{f}$  are coordinates of the potential barrier and the stable equilibrium. $E_{b}$ denotes the height of the potential barrier and $Q$ is the exotermic effect of the transition (energy difference between the metastable and the stable phases).
We consider the propagating transition from the metastable to the stable state. The transition is initiated by forced 'dragging' of the particle at left boundary to a vicinity of the stable state. Then, the conservative dynamics of the system has been simulated under conditions of the free boundaries. Transition front was initiated and moved from left to right. The results of this numeric simulation for different parameters of the system are presented in Fig. \ref{front1}.

First of all, one can observe that after very brief initial transient, the front achieved the constant velocity. The value of this velocity strongly depends on the exotermic effect $Q$. Equally important, in both plots one can observe that the oscillatory tails are not stationary and have quite complicated structure, different for different values of $Q$. In Fig. \ref{front1}a the amplitude of oscillations immediately after the front is relatively low, and then one observes the regime with higher amplitude. In Fig. \ref{front1}b the oscillations amplitude is rather high immediately after the front and then decreases. In both cases, the lengths of the chain fragments with different oscillatory behavior grow as the front propagates. So, one can immediately see that the observed scenario of the front propagation substantially differs from the "idealized" exact solution with the homogeneous oscillatory tail.

Our approximate approach is based first of all on energy considerations. The motion of the front is stationary, therefore one can describe it as a wave of a constant profile, depending on a single effective coordinate $\zeta=n-Vt$, where $V$ is the front velocity. As the front passes certain site, the transition occurs and amount of energy equal to the exotermic effect $Q$ is released.  The system is conservative, so this energy should be stored in the oscillatory zone after the front. In order to provide the stationary front propagation, the oscillations in this zone should be almost monochromatic, with the phase velocity equal to $V$.

\begin{figure}
\includegraphics [width=60mm] {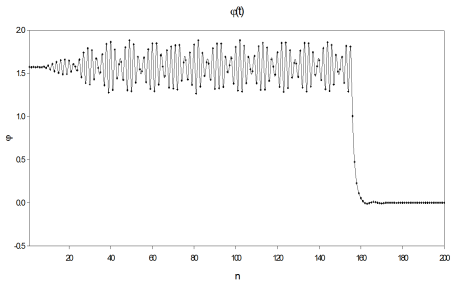}
\caption{An instantaneous displacement profile for the critical wave number $k=\pi$ at the boundary between the first and the second  Brillouin  zones.}
\label{front3}
\end{figure}

In order to estimate the parameters of this process, let us consider the fragment of the chain immediately after the transition front. This fragment contains $L=N-K$ oscillators. Parameter K denotes the artificial left boundary of the oscillatory fragment; the only requirement is that the particle with $n=K$ still belongs to the zone of monochromatic oscillations. Parameter $N$ denotes an instant position of the front and will be defined below in more details. Total energy of this oscillatory fragment is written as:
\begin{equation}
E=\sum_{n=K}^{N} [ \frac{\dot \varphi_{n}^{2}}{2}+\frac{c^{2}}{2}(\varphi_{n+1}-\varphi_{n})^{2}+U(\varphi_{n}) ]
\label{E1}
\end{equation}
Let us consider the time evolution of this energy; we adopt that at time instance $t=N/V$ the particle with number $n=N$ will be exactly at the top of the potential barrier:
\begin{equation}
\begin{split}
\frac{d}{dt}E=\frac{d}{dt}\sum_{n=K}^{N} [ \frac{\dot \varphi_{n}^{2}}{2}+\frac{c^{2}}{4}[(\varphi_{n+1}-\varphi_{n})^{2}+ \\(\varphi_{n}-\varphi_{n-1})^{2}]+U(\varphi_{n}) ]
\end{split}
\label{dE1}
\end{equation}
Taking advantage of equations of motion derived from Hamiltonian (\ref{H1})
\begin{equation}
\ddot \varphi_{n}-c^{2}(\varphi_{n+1}-2\varphi_{n}+\varphi_{n-1}+U^{\prime}(\varphi_{n})=0
\label{eqmot}
\end{equation}
one will obtain:
\begin{equation}
\begin{split}
\frac{d}{dt}E=\frac{c^{2}}{2}(\varphi_{N+1}-\varphi_{N})(\dot \varphi_{N+1} +\dot \varphi_{N})+\varepsilon_{N}\frac{N}{dt}- \\
	\frac{c^{2}}{2}(\varphi_{K}-\varphi_{K-1})(\dot \varphi_{K} +\dot \varphi_{K-1})-\varepsilon_{K}\frac{K}{dt}
\end{split}
\label{dE2}
\end{equation}
Here
\begin{equation}
\begin{split}
\varepsilon_{n}= \frac{\dot \varphi_{n}^{2}}{2}+\frac{c^{2}}{4}[(\varphi_{n+1}-\varphi_{n})^{2}+ \\(\varphi_{n}-\varphi_{n-1})^{2}]+U(\varphi_{n}) \\
n=N, K
\end{split}
\label{Eloc}
\end{equation}
The first term in Equation (\ref{dE2}) corresponds to the energy flow through the transition front, the second one – to the motion of the front itself. The third and the fourth terms describe the energy flow through the "back" boundary and the motion of this boundary.
In order to estimate the amplitude of oscillations after the front, we adopt that the left boundary of the oscillatory zone does not move, i.e. in Eq. (\ref{dE2}) we put $dK/dt=0$. Then, we adopt that all oscillators at the interval $(K,N)$ have the same average energy, equal to $< \varepsilon >=a^{2}\omega^{2}/2$. Thus, from Eq. (\ref{dE2}) we get
\begin{equation}
\begin{split}
< \frac{d}{dt}E >=< \varepsilon >\frac{dL}{dt}= \\ \frac{\omega}{2\pi}\int_{0}^{2\pi \/\omega} dt[\frac{c^{2}}{2}(\varphi_{N+1}-\varphi_{N})(\dot \varphi_{N+1} + \\ \dot \varphi_{N})+\varepsilon_{N}\frac{N}{dt}-
	\frac{c^{2}}{2}(\varphi_{K}-\varphi_{K-1})(\dot \varphi_{K} +\dot \varphi_{K-1})]
\end{split}
\label{dE3}
\end{equation}
Here $dL/dt=V$ describes the growth of the considered fragment of the post-front oscillatory zone. Two first terms in (\ref{dE2}) describe the energy inflow in the oscillatory zone through the moving front, and their sum is equal to $QV$. Here we assume that no small-amplitude waves overrun the transition front. This assumption works if the energy effect of the reaction is large enough. The last term in (\ref{dE2}) describes the energy flow through the immobile left boundary at $n=K$. This flow may be evaluated as the average energy of the oscillations multiplied by a group velocity of the monochromatic wave. Consequently, the condition of energy balance in the oscillatory zone yields:
\begin{equation}
<\frac{E_{L}}{dt}>=<\varepsilon>\frac{dL}{dt}=<\varepsilon>V=QV+<\varepsilon>V_{gr}
\label{dE4}
\end{equation}
So, the average energy of the oscillations in the zone after the front is expressed as follows:
\begin{equation}
<\varepsilon>=\frac{V}{V-V_{gr}}Q
\label{Eavr}
\end{equation}
Here $V_{gr}$ denotes the group velocity of the wave.
So, simple considerations of the energy balance allow one to explain how the non-stationary oscillatory tail is formed: the energy is released with the velocity equal to that of the front propagation, and removed with the group velocity of the wave. So, the average energy of the oscillatory tail generally will not be equal to the exotermic effect of the transition – in other terms, this tail plays a role of the "intermediate" dynamic state achieved by the system immediately after the transition. One should mention that Eq. (\ref{Eavr}) does not depend on the specific choice of the double-well potential – in other terms, this relationship should be true for wide class of model potentials.
From Eq. (\ref{Eavr}) we can see that the average energy of the oscillations is equal to $Q$ only if the group velocity is zero. Such oscillatory tail corresponds to a wave number $k=\pi$ , which divides between the first and the second Brillouin zones of the linear chain.  Displacement profile of the chain in this case is presented in Fig. \ref{front3}.
The critical wave number $k=\pi$ corresponds to the following value of the phase velocity:
\[
V_{*}=\frac{\omega(\pi)}{\pi}=\frac{\sqrt{\omega_{1}^{2}+4c^{2}}}{\pi}
\]
If the front velocity $V>V_{*}$,  then the oscillations in  the tail belong to the first Brillouin zone and the group velocity is positive. Therefore, as it follows from (8), the average energy of oscillations in the tail exceeds  the exotermic transition effect $Q$. It means, in turn, that the energy is pumped into the oscillatory tail also through its left boundary. Then, if this scenario is realized, the balance of the total energy in the chain dictates the cleavage of the oscillatory zone. In order to sustain the growing oscillatory tail, the chain should relax towards the stable stationary state, it is necessary to have a relaxation to the stable stationary state far from the front. Thus, we conclude that the zone after the front in this case should be divided into the "oscillation" and the "relaxation" regions. Exactly this qualitative picture is observed in Fig. \ref{front1}b.

If the front velocity $V<V_{*} $, then the oscillations after the front should belong to the second Brillouin zone. Thus, the group velocity of the wave is negative, and according to (\ref{Eavr}), the average energy of the oscillations after the front will be lower than $Q$. It means that the excessive energy will be pumped to the left end of the chain, and the zone with relatively strong oscillations will be formed after the initial oscillatory tail. In this case, the system will not relax to the stable equilibrium in any of its parts. This generic scenario coincides with the observations presented in Fig. \ref{front1}a.

One can see that the complicated structure of the oscillatory tails after the transition front may be explained by simple combination of the energy considerations and a linear wave mechanics. Moreover, it is possible to predict that two substantially different structural patterns of the post-front oscillations are realized for the cases of small and large values of the exotermic effect $Q$, or for relatively small and large velocities of the front propagation respectively.

From Eq. (\ref{Eavr}) it is easy to obtain the estimation for the amplitude of oscillations, if the latter are considered as approximately linear:

\[
 a=\frac{1}{\omega} \sqrt{2<\varepsilon>}
\]
However, the value of the front velocity (and, consequently, the values of the group velocity and the frequency) remains undefined. In order to determine it approximately, we come back to Eq. (\ref{dE1}) and average it over the time necessary for the transition front to pass exactly one site of the system. General average in flow of the energy through the front over this time should be equal to $Q$. Mathematically, this relationship is expressed as follows:
 \begin{equation}
\begin{split}
\delta E=Q=\int_{(N-1)/V}^{N/V} dt[c^{2}(\varphi(N+1)-\varphi(N)) \\
(\dot \varphi(N+1)+ \dot \varphi(N))+\varepsilon_{N}\frac{dN}{dt}]
\end{split}
\label{deltaE}
\end{equation}

\begin{figure}
a \includegraphics [width=50mm] {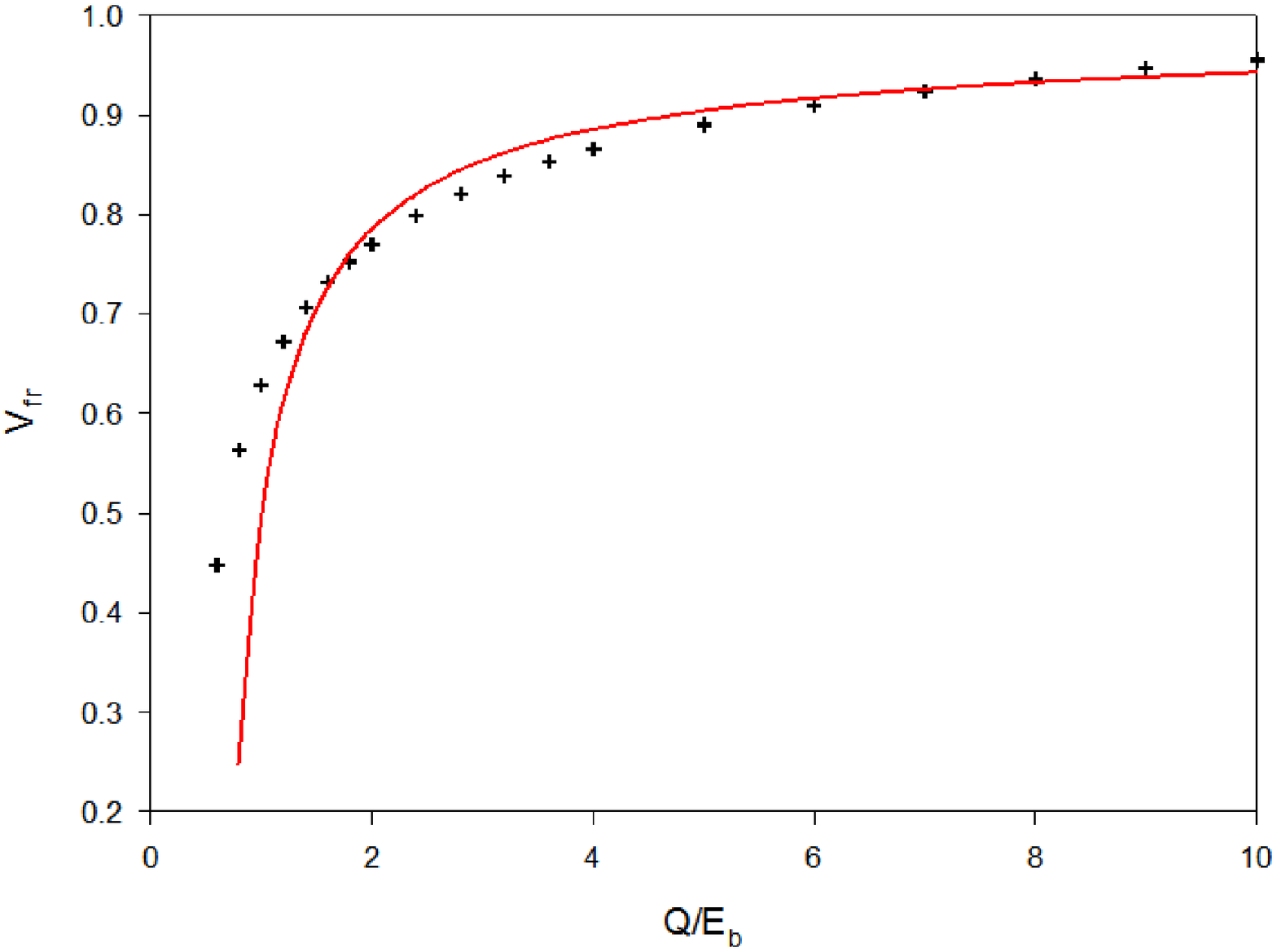} \\ 
b \includegraphics [width=50mm] {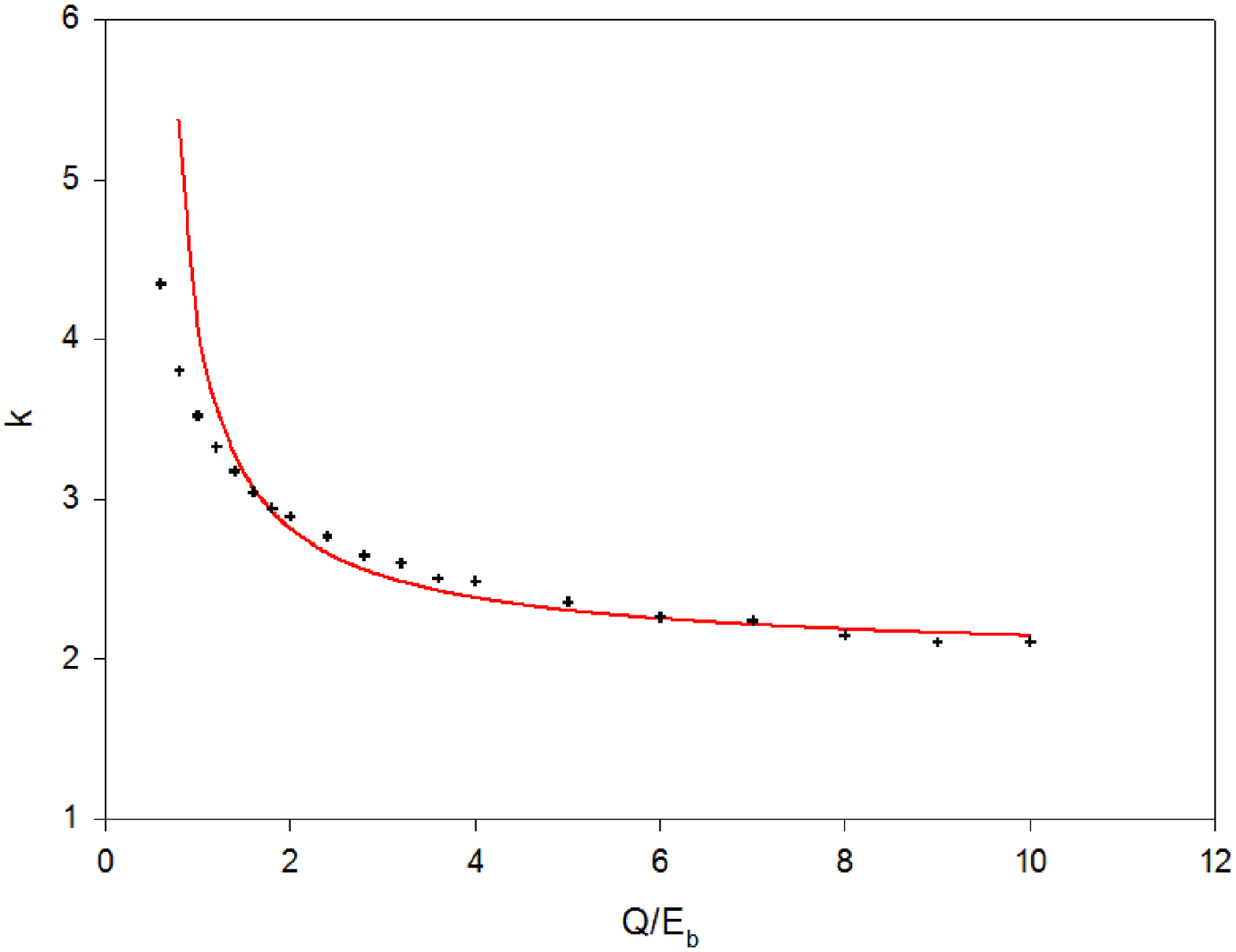} \\ 
c \includegraphics [width=50mm] {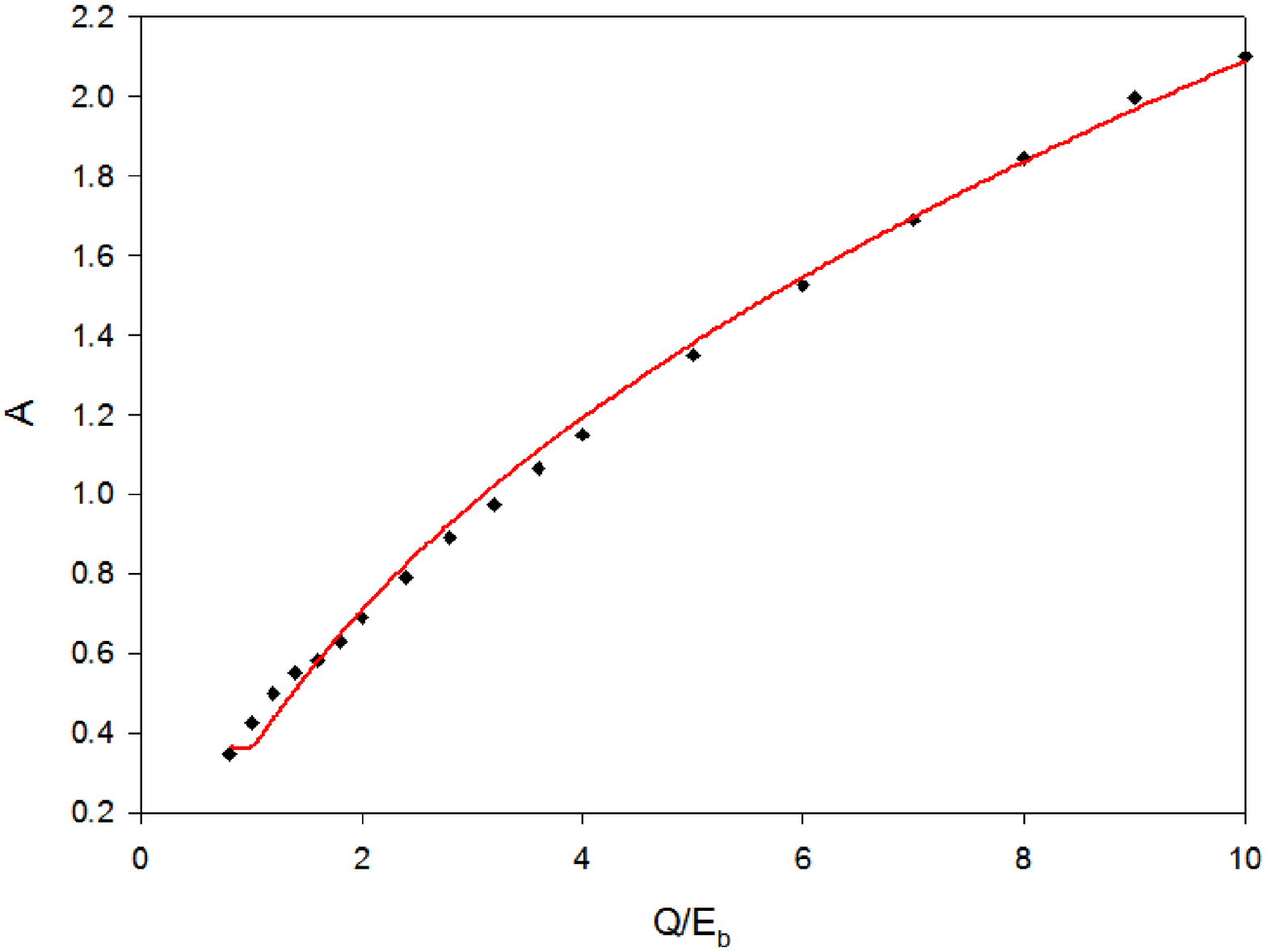}
\caption{Comparison of predictions for (a) kink velocity, (b) wavevector and (c) amplitude of primary oscillatory tail. Solid line represents analytic results, dots - numeric simulation}
\label{comparison}
\end{figure}

For the zone of the chain before the front, one can adopt the following exponential approximation:
\begin{equation}
\varphi(n-Vt)=\varphi_{b} \exp[-(n-Vt)/w]
\label{exp}
\end{equation}
Here $\varphi_{b}$ is equal to the coordinate of the potential barrier , and the value $w$ characterizes the front width. Expression (\ref{exp}) is exactly correct for the model piecewise parabolic potential (\ref{onsite}) in the time instance when the particle is exactly at the top of the potential barrier. For all other time instances and for other model potentials we use (\ref{exp}) as  approximation. Such approximation may be substantiated by the fact that for small displacements the system is approximately linear and thus the exponential approximation is a correct asymptotic solution.  Adopting that the front shape repeats itself after discrete time units $1/V$, we substitute the continuous derivative in \ref{eqmot} by finite difference and arrive to the following approximate relationship between the front velocity and width:
\begin{equation}
2(V^{2}-c^{2})(1-\cosh(\frac{1}{w}))+\omega_{0}^{2}=0
\label{dispersion}
\end{equation}
Substituting (\ref{exp}) into (\ref{deltaE}) and taking into account (\ref{dispersion}), we obtain:
 \begin{equation}
\begin{split}
\frac{Q}{E_{b}}=\frac{e^{-1/2w}\cosh(1/2w)}{w \sinh(1/2w)}[(1-4w^{2}\sinh^{2}(\frac{1}{2w}+ \\
2w^{2}(\cosh(\frac{1}{w})(\cosh(\frac{1}{w})-1)-\\ 4\frac{c^{2}}{\omega^{2}}(1+w(w-1+e^{-2/w}))]
\end{split}
\label{trans}
\end{equation}
The front width $w$ thus can be obtained by solving transcendent equation (\ref{trans}). All other important characteristics of the process, including the velocity of the front propagation, may be computed from (\ref{dispersion}) and from the dispersion relation of the linear chain.
The results delivered by the approximate analytic approach described above are compared with the results of numeric simulations. The results of this comparison are presented in Fig. \ref{comparison}; one can see that the coincidence is rather satisfactory, especially for the most interesting case of not too small values of $Q$.

One can see that considerations based on energy balance and wave mechanics allow rather accurate description of the front propagation and of the oscillatory tail structure. In this Letter we restrict ourselves by very simple benchmark model. Similar treatment should be possible for more complicated models involving gradient nonlinearity and wide range of possible on-site potentials; these issues will be addressed in future research.

\begin{acknowledgments}
The work was supported by Program of Department of Chemistry and Material Science (Program \#1), Russia Academy of Sciences, and Russia Basic Research Foundation (grant 08-03-00420a) and Israel Science Foundation (grant 838/13).
\end{acknowledgments}

\bibliography{CNT_collapse}

\end{document}